\documentclass[12pt,preprint]{aastex}
\usepackage{natbib}
\usepackage{graphicx}
\usepackage{wasysym}
\usepackage{txfonts}
\def\fH2{\mbox {f$_{{\rm H}_2}$}}

\def\EBV{E(B-V)}

\def\nH2{\mbox{${\rm n}(\HH$)}}
\def\enH2{\mbox{$n_{(\HH$)}}}

\def\pccc{~{\rm cm}^{-3}} 
 
\def\pcc {\mbox{${~{\rm cm}^{-2}}$}}

\def\Tsub#1 {\mbox{${\rm T}_{\rm #1}$}}
\def\TK  {\Tsub K }
\def\TB  {\Tsub B }

 \def\arcmin{\mbox{$^{\prime}$}}

\def\degr{$^{\rm o}$}
\def\p{\mbox{$^+$}}

\def\h13cop{\mbox{{H$^{13}$CO\p}}}

\def\c3h2{\mbox{C$_3$H$_2$}}

 \def\R0{R$_0$}
\def\G0{\mbox{G$_0$}}

\def\ddeg{{}^\circ\kern-.1em}

\def\kms{\mbox{km\,s$^{-1}$}}

\def\E#1 {$10^{#1}$}
\def\E#1 {E{#1}}
\def\P#1,{$\nH2\TK~=~#1\times~10^4\pccc$~K}
\def\ec#1,#2,#3,{#1\,(#2)\E{#3}}

\def\H3{\mbox{H$_3$}}
\def\Lya{\mbox{Ly-$\alpha$}}

\def\RH2{\mbox{R$_{\rm G}$}}
\def\g13{\mbox{g$_{13}$}}

\def\kHeH2{\mbox{$k_{ He-\HH}$}}

\def\tim#1,#2{\mbox{{$#1\times10^{#2}$}}}

\newcommand{\emm}[1]{\ensuremath{#1}}   
\newcommand{\emr}[1]{\emm{\mathrm{#1}}} 

\newcommand{\HH}{\emr{H_2}}

\slugcomment{generated \today}

\shorttitle{Red Dawn for H I}
\shortauthors{Harvey Liszt}

\begin{document}


\title{ N(H I)/\EBV}


\author{Harvey Liszt}
\affil{National Radio Astronomy Observatory \\
        520 Edgemont Road, Charlottesville, VA 22903-2475}

\email{hliszt@nrao.edu}





\begin{abstract}

We explore the relationship between dust-emission derived reddening \EBV\ 
and atomic hydrogen column density N(H I) derived from 21 cm emission surveys. 
We consider measurements at galactic latitude $|b| \ga 20$\degr\ and \EBV\ $\la$ 
0.1 mag where the interstellar gas is predominantly  neutral  and atomic and opacity 
corrections to 21cm H I profiles are small.  Over the Galaxy at large at lower 
resolution in H I, and on smaller scales at higher resolution, 
we find that the reddening is always much smaller than would be expected 
from the usually-quoted relation N(H) $ = 5.8\times 10^{21}\pcc$ 
\EBV\ based on stellar reddening and {\it uv } absorption toward early-type stars. 
On wide scales, we find 
N(H I) = $8.3 \times 10^{21}\pcc$ \EBV.  We cite various precedents for
such a large N(H I)/\EBV\ ratio whenever wide-field 21cm emission surveys are
considered, including when reddening based on galaxy counts or colors is substituted 
for the dust-emission derived reddening measure.

\end{abstract}



\section{Introduction}

A linear relationship between optical reddening \EBV\ and the total column density 
of hydrogen nuclei N(H) = N(H I) + 2 N(\HH) is a cornerstone of modern astrophysics,
with a widely-quoted constant of proportionality 
N(H) $= 5.8 \times 10^{21}\pcc$ \EBV\ derived from \Lya\ and \HH\ absorption 
against a small number of early-type stars using the {\it Copernicus} satellite 
\citep{BohSav+78,SavDra+77}.  The same constant of proportionality is implicit 
in the definition of the ``Standard`` H I cloud of \cite{Spi68}.

This relationship was at first extended over the sky using galaxy counts to 
measure optical extinction \citep{BurHei82} (BH82; see also \cite{Hei76}) 
and more recently using far infrared (FIR) dust emission-derived measures 
from higher-resolution (6\arcmin) all-sky surveys \citep{SchFin+98} (SFD98).  
For the atomic hydrogen column density, the all-sky, stray radiation-
corrected LAB 21cm H I emission survey \citep{KalBur+05} with 0.6\degr\
resolution on a 0.5\degr\ grid stands in for \Lya\ absorption.  MM-wave 
CO emission surveys \citep{DamHar+01,AbdAck+10,AdeAgh+11} have been used as 
surrogates for \HH\ at higher reddening \EBV\ $>$ 0.1 mag where the \HH\
fraction is appreciable \citep{SavDra+77} and N(H) necessarily diverges 
from N(H I).  The \HH-fraction has also been estimated using 2N(\HH) = 
$5.8 \times 10^{21}\pcc$ \EBV\ - N(H I) \citep{AbdAck+10,LisPet+10,LisLuc02}, 
partly as a means of calibrating the so-called CO-\HH\ conversion factor.

\begin{figure*}
\includegraphics[height=7cm]{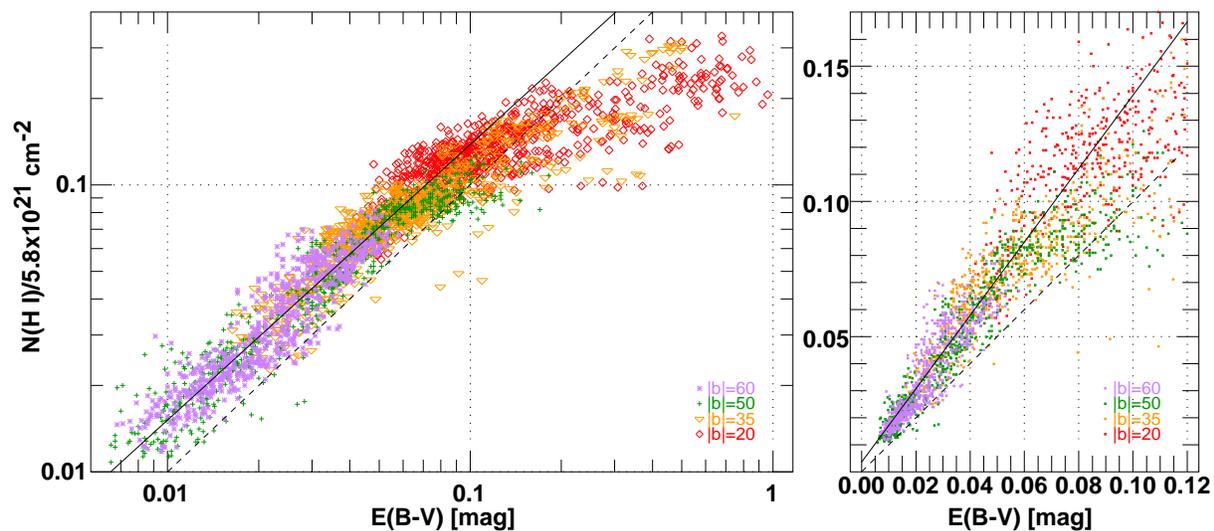}
 \caption[] {Reddening \EBV\ from SFD98 and scaled LAB-survey H I column density 
N(H I)/$5.8\times10^{21}\pcc$.  N(H I) was derived assuming a spin temperature of 
145 K to correct for optical depth.  N(H I) and \EBV\ were sampled at the grid points 
of the LAB survey, ie at even 0.5\degr\ intervals around the sky at
galactic latitudes b = $\pm$20\degr, $\pm$35\degr, $\pm$50\degr\ and $\pm$60\degr.  
The dashed line in each panel represents the canonical scaling of \EBV\ and N(H), 
the solid line is a regression fit.  A handful of points at \EBV\ $>$ 1 mag
have been omitted in the left-hand panel.  The right hand panel shows the
data on a linear scale at smaller values of \EBV.}
\end{figure*}


The subject of this work is the relationship between N(H I) derived from 21cm
measurements and \EBV\ at 0.02 mag $<$ \EBV\ $<$ 0.1 mag.  In this regime the 
\HH-fraction is small, the hydrogen is predominantly  neutral and atomic and the 
optical depth in the 21 cm H I line is low.  Thus all of the commonly-considered 
systematic errors in deriving N(H I) are minimized and there should be little 
difficulty in demonstrating the proportionality between 
N(H I) {\it cum} N(H) and \EBV.  However, we find that the slope of the 
N(H I)-\EBV\ relationship is very different from that expected, 
typically by about 40\%, over large and small regions of the sky, at 
low and high spatial resolution, no matter which recent H I survey is 
considered.

\begin{figure}
\includegraphics[height=8.4cm]{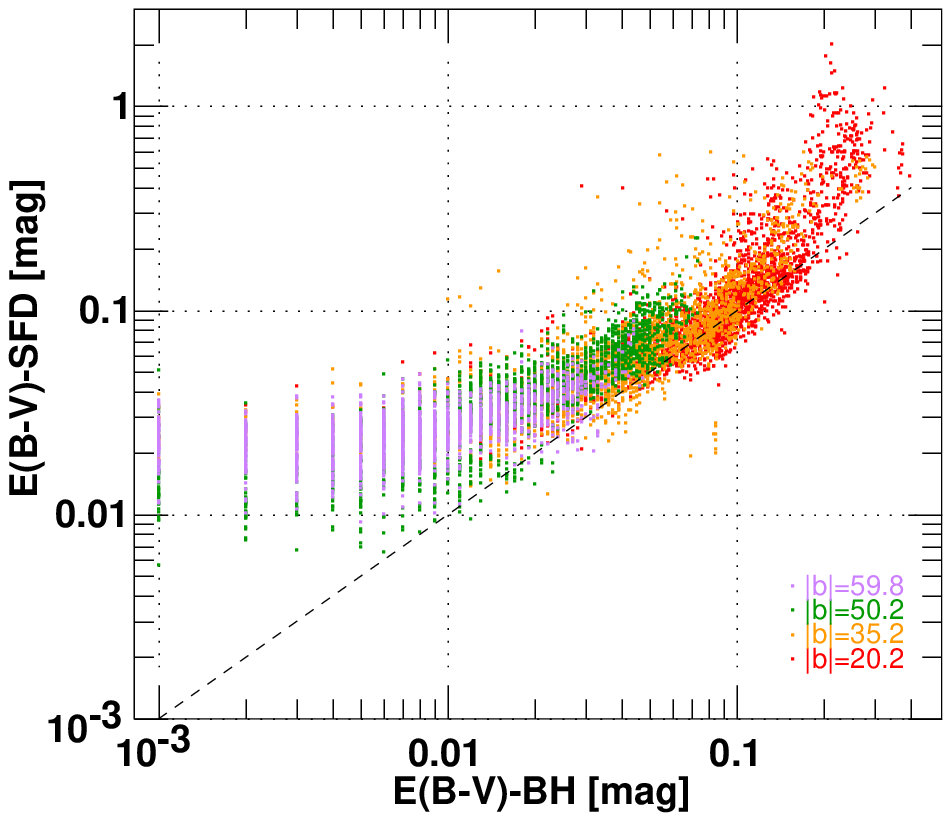}
 \caption[]{Reddening \EBV\ derived from the datasets of SFD98 andBH82
along tracks that most closely approximate those used to create  Figure 1.
The reddening was sampled at all 1200 longitude grid-points of the BH82
dataset at positive and negative values of the latitudes indicated. } 
\end{figure}

The organization of this work is as follows.  In Section 2 we describe
our use of the various H I and reddening measures and databases.  In 
Section 3 we show the N(H I)-\EBV\ relationships that result.  In Section 
4 we discuss some recent and historical antecedents for adjusting the
\EBV/N(H I) relationship and Figure 5 is a brief summary presenting
the alternatives that suggest themselves for understanding the present
results.

\section{Observational material and empirical results}

\subsection{Reddening}

For comparison with N(H I) we derived reddening values at the nearest 
pixel in the SFD98 datacubes (North and South) and made sure to check 
our values against those served 
online \footnote{http://irsa.ipac.caltech.edu/applications/DUST/}
given the unexpected nature of the plots we produced.  The SFD98 datacubes
present equivalents of the optical reddening \EBV\ derived from far 
infrared dust emission at 6\arcmin\ resolution on a 2.5\arcmin\ pixel grid,
 with a stated error of 16\% at each pixel.  We also compared the SFD98 results 
with those from the work of BH82, on the gridpoints of the latter, 
ie spaced 0.6\degr\ in galactic latitude and 0.3\degr\ in longitude.

\subsection{21cm H I}

We derived N(H I) from brightness temperature profiles of 21cm radiation, 
lightly correcting for optical depth using a spin temperature of 145 K, which 
corresponds to the mean ratio of 21cm emission and absorption 
\citep{LisPet+10} in the Millenium H I Emission-Absorption Survey of 
\cite{HeiTro03}.  The correction to N(H I) for H I optical depth is never more 
than a few percent at N(H I) $< 5\times 10^{20}\pcc$ , \EBV\ $<$ 0.08 mag where 
the disparity between the canonically-scaled \EBV\ and N(H I) is most clearly manifest.
The upward correction for the possible contribution of ionized hydrogen is
ignored here, but it would only exaggerate the effects that are discussed. 

On large scales we used the LAB all-sky, stray-radiation corrected survey data
\citep{KalBur+05} that has 0.6\degr\ spatial resolution on a pixel grid spaced 
0.5\degr, with a velocity resolution of 1.3\kms.  We checked our profile integrals 
against those from the online LAB survey 
server \footnote{http://www.astro.uni-bonn.de/hisurvey/profile/}.

We also derived N(H I) over smaller regions using cubes of H I profiles from the 
GALFA-HI survey \citep{PeeHei+11}\footnote{https://purcell.ssl.berkeley.edu/}
that has 4\arcmin\ resolution on a 1\arcmin\ pixel grid with 0.8 \kms\ velocity 
resolution. We confirmed that the GALFA and LAB surveys are on the same intensity 
scale by synthesizing a few spectra at 0.6\degr\ resolution at positions 
corresponding to LAB survey pixels, using GALFA data.

\section{Empirical results}

\begin{figure}
\includegraphics[height=8.4cm]{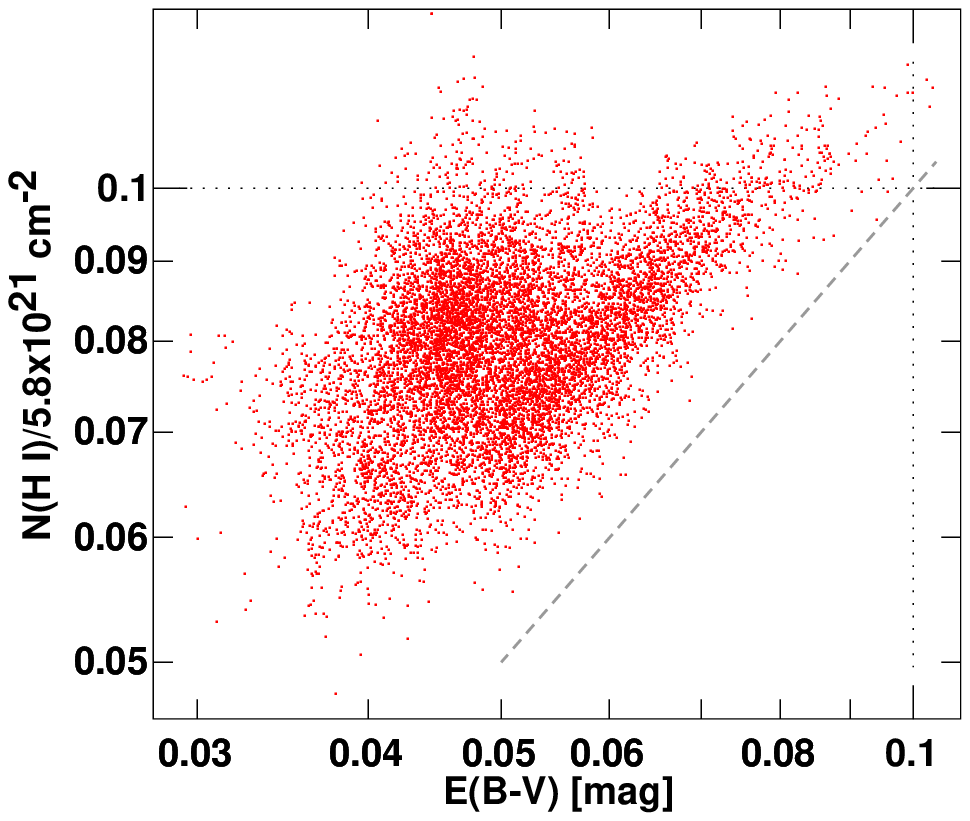}
 \caption[]{Reddening from SFD98 and N(H I) from the GALFA survey 
at 4\arcmin\ resolution corrected for a spin temperature of 145 K,
 in a 5\degr\ field centered at RA(J2000)=  18h30m, Dec(J2000) = 
22\degr30\arcmin\ (l = 82.099\degr, b = -32.931\degr) chosen to 
avoid sightlines at \EBV $>$ 0.1 mag. The dashed line is the canonical 
scaling  N(H I) =  $5.8 \times 10^{21}\pcc$ \EBV, as in Figure 1.}
\end{figure}

\subsection{Wide-angle results using the LAB H I survey}

Figure 1 shows the results of plotting scaled values of 
N(H I)/$5.8\times10^{21}\pcc$ vs \EBV\ from SFD98.  We sampled around the entire 
sky at the 720 longitude gridpoints of the LAB survey, ie at even 0.5\degr\ intervals 
of galactic longitude at b = $\pm$20\degr, $\pm$35\degr, $\pm$50\degr\ 
and $\pm$60\degr, to cover a wide range in N(H I) and \EBV;  we used the
nearest datapoint in the 2.5\arcmin-pixel datacubes of SFD98.  
Figure 1 shows the data on a wider log scale at left and at smaller \EBV\ on a 
linear scale at right.

The dashed lines in Figure 1 are the canonical scaling
relationship N(H I)/$5.8\times10^{21}\pcc$ = \EBV\ which should hold at
lower \EBV\ where the \HH\ fraction is small \citep{SavDra+77}.
A regression fit to the data at 0.015 $\le$ \EBV\ $\le$ 0.07 mag is  
log(y) = $(0.100 \pm 0.016) + (0.960\pm 0.011)$ log(\EBV) where 
y = N(H I)/$5.8\times10^{21}\pcc$.

As shown at right there is a small non-zero y-intercept to a linear fit, 
y = 0.00363+1.3581 \EBV\ or 
\EBV\ = -0.00267+N(H I)/$7.87\times 10^{21}\pcc$ which is very similar to
an implicit result in \cite{Pee13}, as noted here in Section 4.2.

A formal fit for the mean displacement between the regression lines and the
canonical scaling is a factor 1.43, ie a mean slope dN(H I)/d\EBV\ = 
$8.3\times10^{21}\pcc$ \EBV.  Note that the putative displacement cannot 
be discerned at \EBV\ $>$ 0.12 mag, in part because the \HH\ fraction is 
appreciable and partly because the optical depth at 21cm is not compensated 
in the attendant cooler gas.

As a check, Figure 2 shows that the reddening derived from the SFD98 
datacubes is already larger than that of BH82, except over a narrow range 
at 0.06 $<$ \EBV\ $<$ 0.1 mag.  Recent suggestions for corrections to
the SFD98 results 
 \citep{SchFin+10,SchFin11,BerIve+12} imply an overall decrease
of 10-15\% in the SFD98 reddening that would somewhat ameliorate 
the differences between SFD98 and BH82 while increasing the value we
derive for N(H I)/\EBV\ by the same amount.

Given the behaviour in Figure 2 it is interesting to contemplate
what Figure 1 would have looked like if had been constructed just before
the appearance of SFD98.  The reddenings of BH82 do not increase much
beyond \EBV\ = 0.2 mag, so that the broad wing extending to high
\EBV\ in Figure 1 would have been compressed and pushed back to the 
regression line. At small \EBV\ the reddening of BH82 declines much faster 
than that of SFD98, leading to high values of the gas/dust ratio as noted 
in Section 4.2.

Table 1 shows some summary properties for the data displayed in Figure 1.
The southern sky consistently has 20-30\% more gas at $|b| \ge 35$\degr, 
a well-known aspect of galactic structure \citep{DicLoc90}.  The 
H I/\EBV\ ratio is more consistent in the north but steadily increasing
with $|b|$ in the south.  Some part of that increase could be explained by
colder gas and a stronger correction for saturation in the H I but it also
appears that the extra gas in the south is not accompanied by a full
complement of added reddening.  In any case, Table 1 gives some idea of
the expected variability in N(H)/\EBV\ at/above 35\degr\ where the 
opacity correction is small; nearly none in the northern sky and 
15-20

\subsection{Smaller-scale results using GALFA HI}

\begin{figure*}
\includegraphics[height=7cm]{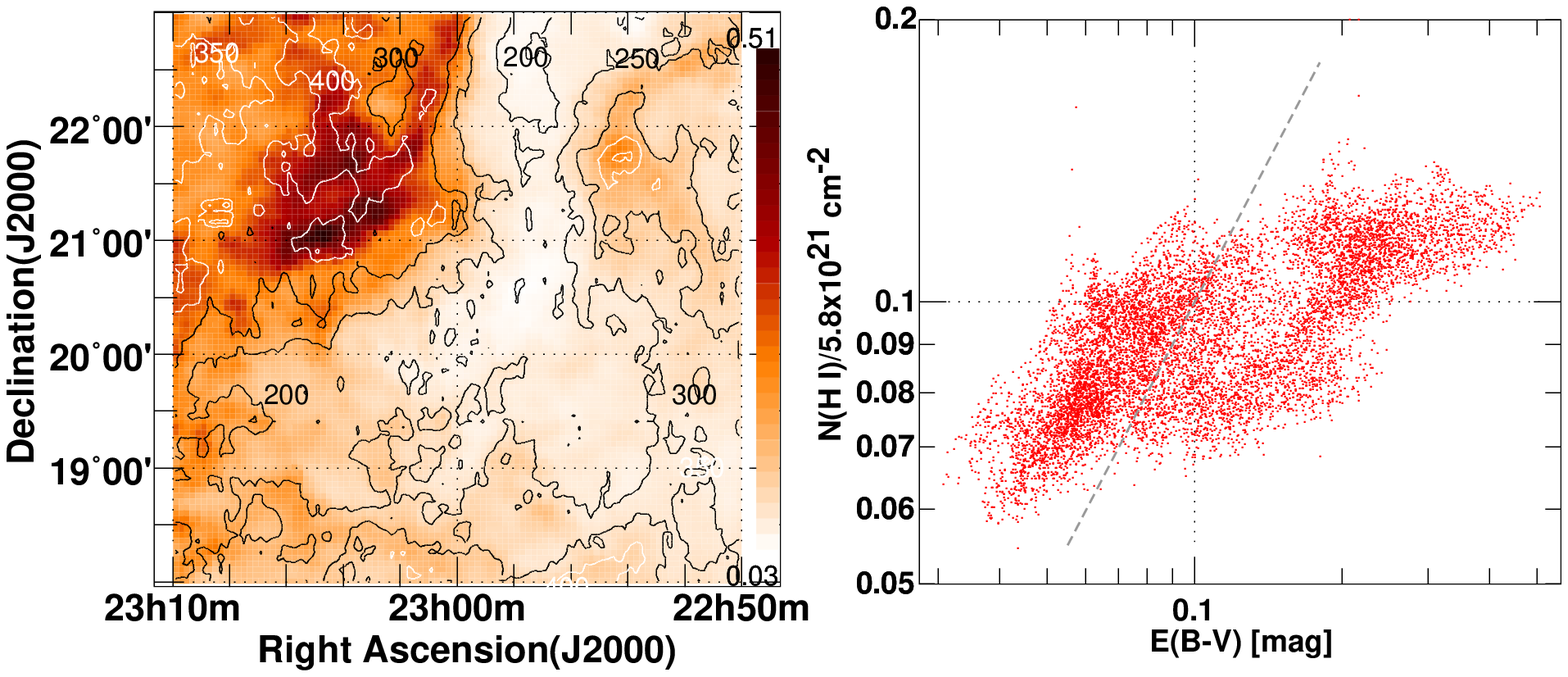}
 \caption[]{Reddening and H I in a 5\degr\ field centered at RA(J2000) = 23h,
Dec(J2000) = 20\degr30\arcmin, near the high-latitude molecular cloud MBM53.
Left: Pseudo-color map of reddening \EBV\ (0.03 $<$ \EBV\ $<$ 0.51 mag) 
with overlaid contours of the H I profile integral (units of K-\kms) 
using 4\arcmin-resolution HI data from the GALFA survey.  
Right: plot of N(H I) corrected for optical depth using a spin temperature 
of 145 K vs. reddening.  The dashed line is the canonical 
scaling  N(H I) =  $5.8 \times 10^{21}\pcc$ \EBV, as in Figure 1.}
\end{figure*}

As a check on the wide-scale results from the LAB survey,  we considered 
regions mapped at 4\arcmin\ resolution in  the GALFA-HI survey
of \cite{PeeHei+11}.  Shown in Figures 3 and 4 are results derived from two custom 
5\degr$\times$5\degr\ cubes delivered by the GALFA data server.
 
Figure 3 shows the N(H I) - \EBV\ comparison for a region at b $\approx 
$-33\degr\ that was specifically selected from a reddening map to avoid sightlines 
with \EBV\ $\ga$ 0.1.  The formal value of the displacement of the data from the 
canonical scaling is a factor 1.53, slightly larger than in Figure 1.

Shown at left in Figure 4 are overlays of the reddening (in pseudo-color) and
contours of the H I profile integral for a region specifically chosen to contain 
a wide range in reddening and portions of the high-latitude CO-emitting cloud MBM53 
\citep{MBM85}.  This part of the sky near the MBM53-55 complex harbors a relatively 
large \HH-fraction even in regions where \EBV\ $\la 0.1$ mag \citep{LisPet12}. 
Although the \EBV\ map and the contours of the H I profile integral correspond
quite closely, the plot at right in Fig. 4  makes clear that the range of the 
H I column density is very much compressed compared to that of the reddening.  
The H I profiles are depressed by the presence of cold gas and a high molecular 
fraction at relatively small \EBV\  but N(H I) still lies well above the dashed 
line representing the canonical scaling between N(H) and \EBV\ at smaller values of 
\EBV.

\section{Discussion}

\begin{figure}
\includegraphics[height=8.4cm]{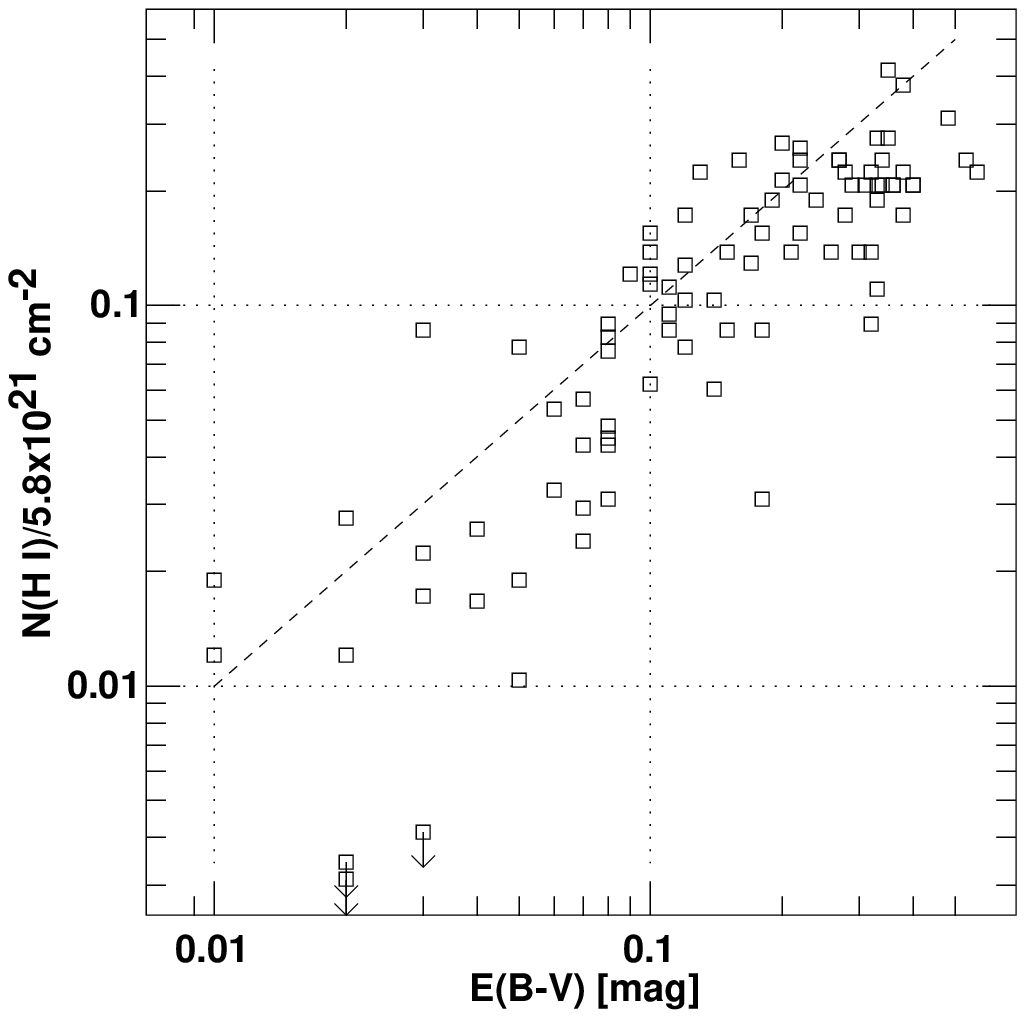}
 \caption[]{Stellar reddening and N(H I) derived from \Lya\ measurements
by \cite{BohSav+78}.} 
\end{figure}

\subsection{Suggested corrections to SFD98}


We noted in  Section 3.1 and Figure 2 that the SFD98 results are generally larger 
than those of BH82.  \cite{SchFin11} and \cite{SchFin+10} derived a 14\% downward
overall scaling of the SFD98 results and they and \cite{BerIve+12} discuss corrections to the 
SFD98 reddening as well as a normalization difference between the north and south 
hemispheres of about 20\%.  Downward scaling of the SFD98 results would
lessen some of the differences seen in Figure 2 while exacerbating the disparity in the 
 \EBV-N(H I) comparison discussed here.  \cite{PeeGra10} argued 
that the SFD98 reddening could be low by as much as 50\% over limited portions of 
the sky where the dust temperature is low but generally confirmed the SFD98
values otherwise.

\subsection{Prior uses of 21cm data}

With respect to the radio data, the effect manifested in Figures 1, 3 and 4 is not 
subtle and it is perhaps surprising that it has not been previously called out .  
However, the effect can be found in the literature in various ways when it is sought;
for instance Figure 4 of \cite{AbdAck+10} where the SFD98 reddening is uniformly
too small outside the molecular regions, compared to the canonically-scaled N(H I).
This is as expected from the plot in Figure 1, showing the suppression of the 
H I profile integral for \EBV\ $>$ 0.1 mag.

\cite{Pee13} related both \EBV\ and a scaled representation of the optically 
thin values of N(H I) to his galaxy-color-derived reddening measure over a wide
region around the North Galactic Cap.  The implied \EBV-N(H I) relationship from 
his equations 1 and 2 is \EBV\ $= -0.00243+$N(H I)$/7.64\times10^{21}\pcc$, very
similar to the linear fit to the data shown in Figure 1, as 
noted in Section 3.1.  At \EBV\ = 0.06 mag
one has N(H I)/\EBV\ = $7.95\times 10^{21}\pcc$/mag,  tantamount to a 
37\% rescaling of the H I-\EBV\ ratio, compared to our fitted value of 43\% using 
(lightly) optical-depth corrected N(H I) in Figure 1.

In the original discussion of \cite{BohSav+78} it was noted that the slope of the
N(H I)-\EBV\ relationship derived from contemporaneous 21cm H I measurements and
galaxy counts was somewhat smaller, dN(H I)/d\EBV\ = 
$4.85 \times 10^{21}\pcc$ mag$^{-1}$ in \cite{Hei76}.  This was consistent 
with the comparison between \EBV\ and N(H I) derived
from the \Lya\ line, see just below.  However, although the slope 
was correctly quoted, the discussion of \cite{BohSav+78} glosses over an
important aspect of  the regression fit derived by \cite{Hei76}, 
\EBV\ = $(-0.041\pm0.012)$ + N(H I)/$(4.85\pm 0.36) \times 10^{21}\pcc$.
Regardless of whether the negative intercept is physically-based, as
\cite{Hei76} suggested might be the case (ie, gas without dust),
it reflects very large ratios N(H I)/\EBV\ at low \EBV,  for instance
N(H I)/\EBV\ = $8.2\times10^{21}\pcc$ mag$^{-1}$ at \EBV\ = 0.06 mag,
quite in line with the present discussion.

\subsection{N(H I) derived from \Lya\ absorption}

The N(H I) data of \cite{BohSav+78} derived from \Lya\ absorption 
are shown in Figure 5 and the behaviour is mostly opposite to that displayed 
in Figure 1 here.  At lower \EBV\ the scatter is larger and the H I column 
densities are below $5.8 \times 10^{21}\pcc$ \EBV.  

It was well understood that the sightlines selected for {\it uv} absorption 
measurements by {\it Copernicus} had low mean density and were biased toward 
lower \HH-fractions \citep{BohSav+78}.  This in part explains the near 
coincidence of N(H I) and the canonical scaling of \EBV\ at \EBV $>$ 0.1 mag 
in Figure 5,
well beyond the point at which \EBV\ and N(H I) diverge in Figure 1.  However, 
assuming a fixed gas/dust ratio, there would be no reason to correct the 
derived N(H)/\EBV\ for a small \HH\ fraction if both N(H I) and N(\HH) were 
accurately measured, no matter the sample biases. 

The somewhat small values of N(H I)/\EBV\ at \EBV\ $< 0.1$ mag 
 in Figure 5 could be 
explained by adding a 15\% contribution for hydrogen that is ionized 
along the sightlines to those early-type stars:  \cite{BohSav+78} quote 
N(H I)/\EBV $= 5\times 10^{21}\pcc$ mag$^{-1}$).  However, aligning the 
radio and {\it uv} results below \EBV\ = 0.1 mag in this same way
would require a very large contribution from ionized gas, of order 60\%
of the observed N(H I) derived from \Lya.  Alternatively, the grains observed 
toward these stars may actually be different.

 \cite{DipSav94} (see also \cite{KimMar96}) used archival IUE spectra and 
expanded the sample of N(H I) derived from \Lya\ absorption by about a factor 5. 
They quote N(H I)/\EBV $= 4.93\times 10^{21}\pcc$ mag$^{-1}$ at 
\EBV $>$ 0.1 mag, just as in \cite{BohSav+78}.  However, for \EBV $<$ 0.1 mag 
their measurements give much higher N(H I)/\EBV, more in line with what we 
derive, and opposite to the data of \cite{BohSav+78} shown here in Figure 5.

\section{Summary and conclusions}

We compared IR-derived measures of \EBV\ and N(H I) derived from wide-field
H I sky surveys at low (36\arcmin; Figure 1) and high (4\arcmin, Figures 3 and 4) 
spatial resolution in the 21 cm line and found a 40\%-50\% displacement between 
the resultant N(H I)/\EBV\ ratio $\approx 8.3 \times 10^{21}\pcc$ mag$^{-1}$ and 
the usually-quoted scaling 
N(H) $= 5.8 \times 10^{21}\pcc$ \EBV, at least at \EBV\ $\la$ 0.1 mag
where the \HH\ fraction is low and little saturation of the  H I profiles is
evident.  In any case, the commonly-cited systematic errors in deriving N(H I) from
the H I profile, and the systematic deviations of N(H) and N(H I) to account
for the contribution of ionized gas, would all increase the hydrogen column density 
and exaggerate the disparity between the two scaling relations.

Along the way we digressed to note that the IR dust emission-derived reddening 
is already larger than that derived from earlier all sky surveys of galaxy 
counts and that recent suggestions for recalibration of the reddening would 
lessen the disparity with earlier studies of galaxy counts but widen it with 
respect to our H I measure.  We showed that there have been intimations of a 
high 21 cm-N(H I)/\EBV\ ratio for \EBV\ $<$ 0.1 mag for some time but that 
a somewhat opposite effect was seen in {\it uv} \Lya\ measurements of N(H I)
 toward early-type stars.

There is no doubt that the gas column density and dust optical depth are well 
and, apparently, linearly correlated on large scales at \EBV\ $\la$ 0.1 mag, 
but the coefficient of proportionality seems to have been mis-estimated
by a surprisingly large amount, for a surprisingly long while.  If the 
FIR-derived \EBV\ and 21 cm-derived N(H I) have been measured correctly, 
they imply a 30\% decrease in the accepted galactic \EBV/N(H) ratio
at \EBV\ $\la$ 0.08 mag where N(H I) $\simeq$ N(H).  Moreover, the
obvious sources of possible error, opacity in the H I and the unobserved 
contribution of H II, both go in the direction of increasing the disparity.
Conversely, if the stellar reddening and {\it uv} absorption measurements 
of H I and \HH\ are also accurate, either a very large contribution from 
ionized gas was neglected toward the early-type stars or the dust optical
depths per H-nucleus are larger than in the Galaxy at large.

In a subsequent paper we will discuss the physical reasons for the
divergence between N(H I) and the linear \EBV-N(H I) scaling at 
\EBV\ $<$ 0.1 mag.  By 
considering the H I profiles and the established empirical relationship 
between \EBV\ and the integrated H I optical depth, it is straightforward to 
show that saturation of the 21cm line is much less important than the
conversion of hydrogen atoms to \HH.

\acknowledgments

  The National Radio Astronomy Observatory is operated by Associated
  Universities, Inc. under a contract with the National Science Foundation.
  The author was partially funded by the grant ANR-09-BLAN-0231-01 from the 
  French {\it Agence Nationale de la Recherche} as part of the SCHISM 
  project (http://schism.ens.fr/).  The author acknowledges the
  hospitality of Newrest at the ALMA OSF facility at 2800m in the Atacama 
  and the hospitality of the Hotel Director in Santiago during the writing
  of this manuscript.  I thank Bruce Draine for words of encouragement and 
  I thank the referee, Josh Peek, for his helpful remarks.

  This publication utilized Galactic ALFA HI (GALFA HI) survey data 
  obtained with the Arecibo L-band Feed Array (ALFA) on the Arecibo 305m 
  telescope.  The GALFA HI surveys are funded by the NSF through grants to 
  Columbia University, the University of Wisconsin, and the University of 
  California. 




\bibliographystyle{apj}



\begin{table}
\caption[]{Mean quantities for the data displayed in Figure 1}
{
\small
\begin{tabular}{rccc}
\hline
{\it b} &$< \int \TB dv >$ & $<\EBV>$ &  $<\int \TB dv>/<\EBV> ^a $\\
       &  K-\kms & mag & ${\pcc {\rm mag}^{-1}}$ \\
\hline
-60\degr & 124 & 0.0238 & 9.92 $\times 10^{21}$ \\  
60\degr  & 102 & 0.0234 & 8.13  $\times 10^{21}$ \\  
\hline
-50\degr & 178 & 0.0431 & 8.61  $\times 10^{21}$ \\  
50\degr & 136 & 0.0318 & 8.31  $\times 10^{21}$ \\  
\hline
-35\degr & 312 & 0.0951 & 7.36  $\times 10^{21}$ \\  
35\degr & 252 & 0.0633 & 8.07  $\times 10^{21}$ \\  
\hline
-20\degr & 439 & 0.1821 & 6.55  $\times 10^{21}$ \\  
20\degr & 412 & 0.1569 & 7.14  $\times 10^{21}$ \\  
\tableline
\end{tabular}}
\\
$^a$ rescaled by $1.823\times 10^{18}\pcc$ to give optically thin N(H I)\\
\end{table}

\end{document}